\documentclass[aps,twocolumn,notitlepage,superscriptaddress, nofootinbib]{revtex4-2}

\usepackage{xcolor,graphicx}
\usepackage{amssymb,amsmath,graphicx}
\usepackage[utf8]{inputenc}
\usepackage{comment}
\usepackage{braket}
\usepackage[normalem]{ulem}
\usepackage{hyperref,url}
\usepackage{float}

\usepackage{siunitx} 
\newcolumntype{T}{j[table-format=3.5]}
\usepackage{cellspace}

\usepackage{xcolor}
\hypersetup{
    colorlinks,
    linkcolor={red!50!black},
    citecolor={blue!50!black},
    urlcolor={blue!80!black}
}

\usepackage{longtable}

\usepackage{mathtools}

\begin{document}

\title{Tight Bell inequalities from polytope slices}

\author{José Jesus}
\affiliation{Instituto Superior Técnico, 1049-001 Lisbon, Portugal}

\author{Emmanuel Zambrini Cruzeiro}
\affiliation{Instituto de Telecomunica\c c\~oes, 1049-001 Lisbon, Portugal}

\begin{abstract}
We derive new tight bipartite Bell inequalities for various scenarios. A bipartite Bell scenario $(X,Y,A,B)$ is defined by the numbers of settings and outcomes per party, $X,A$ and $Y,B$ for Alice and Bob, respectively. We derive the complete set of facets of the local polytopes of (6,3,2,2), (3,3,3,2), (3,2,3,3), and (2,2,3,5). We provide extensive lists of facets for (2,2,4,4), (3,3,4,2) and (4,3,3,2). For each inequality we compute the maximum quantum violation, the resistance to noise, and the minimal symmetric detection efficiency required to close the detection loophole, for qubits, qutrits and ququarts. Based on these results, we identify scenarios which perform better in terms of visibility, resistance to noise, or both, when compared to CHSH. Such scenarios could find important applications in quantum communication.
\end{abstract}

\maketitle

\section{Introduction}

Bell inequalities are central to the study of non-locality, but finding the complete list of Bell inequalities for a given Bell scenario can be a very difficult task \cite{Pitowsky1989}. A Bell scenario is specified by a number of measurement settings and a number of measurement outcomes for each party. We specify a scenario by giving those numbers $(X,Y,A,B)$, where $X,Y$ are the number of inputs for Alice, Bob, resp. and $A,B$ are the corresponding numbers of outcomes. In the case of two parties with two measurement choices each (the simplest case), $(2,2,2,2)$, there is only one tight Bell inequality, the Clauser-Horne-Shimony-Holt (CHSH) inequality \cite{Clauser1969}. The local polytope has two classes of facets, CHSH and positivity. If one allows both parties to choose between three binary outcome measurements, $(3,3,2,2)$, there is only one new relevant inequality besides CHSH \cite{Froissart1981}. For four settings on each side, $(4,4,2,2)$, the number of classes of facet inequalities grows to 175, where 169 of these inequalities genuinely use the four settings. The complete list of inequalities has not been known until recently \cite{Brunner2008,Pal2010,Bancal2010,Deza2020}. The quantum bound, resistance to noise for partially and maximally entangled states, and detection efficiencies for all these inequalities were then presented in \cite{Cruzeiro2019}, for the case of two-qubit states.

On the other end of the spectrum, for scenarios with a binary choice of measurements and arbitrary numbers of outcomes, little is known. The full list of facets for $(2,2,3,3)$ is known, and consists of only three classes \cite{Collins2002,Kaszlikowski2002,Collins2004}.

We present the complete list of facets, along with their properties, for several scenarios: $(6,3,2,2)$, $(2,2,3,5)$, $(3,2,3,3)$ and $(3,3,3,2)$. In $(6,3,2,2)$, we find 7 classes of facets. This scenario does not yield any new inequality when compared to $(5,3,2,2)$. Scenario $(2,2,3,5)$ does not yield any new facets with respect to $(2,2,3,3)$. Finally, we present all 38 classes of facets for $(3,2,3,3)$ and all 25 classes of facets for $(3,3,3,2)$. We also present (possibly complete) lists of facet classes for $(2,2,4,4)$, $(3,3,4,2)$ and $(4,3,3,2)$. In addition, we study the basic quantum properties of these inequalities by computing the local bound, two-qubit, two-qutrit, and two-ququart, quantum bounds, the concurrence of the state which attains the optimal quantum bound for qubits, the resistance to noise and finally the minimum detector efficiency required to close the detection loophole assuming Alice and Bob's detectors have the same efficiency. The resistance to noise and minimum detection efficiency were computed for the state and measurements that maximally violate the inequality. Furthermore, we reanalyse the facets of $(4,4,2,2)$ for the cases of two-qutrit and two-ququart states. 

In Section \ref{sec:two}, we review the methods used in the present article to analyse and classify facets. In Section \ref{sec:three}, we describe the computation of the local and quantum bounds, the resistance to noise, concurrence, and minimal detection efficiency to close the detection loophole. In Section \ref{sec:four}, we discuss our results regarding the slicing procedure. In Section \ref{sec:five}, we select amongst our facets the most promising candidates for quantum communication. The lists of facets are provided in a supplementary file, while their properties are presented in the Appendices.

\section{Methods} \label{sec:two}

\subsection{Convex polytopes}

A \textit{convex polytope} can be defined in a number of ways \cite{Ziegler2012}. Two important definitions are: 1) the convex hull of a set of points, 2) the intersection of a set of half-spaces \footnote{A \textit{half-space} is any of the two parts of an affine space cut in two by a hyperplane.}.

For a finite set of points $K = \{x_1,\dots,x_n\}\in\mathbb{R}^d$, the convex hull is
\begin{equation}
 \mathrm{conv}(K) = \{ \sum_{j=1}^n \lambda_jx_j\ :\ \lambda_j\geq 0,\ \sum_j \lambda_j = 1\}
\end{equation}
while the intersection of a number of half-spaces is
\begin{equation}
 P(A,z) = \{x\in\mathbb{R}^d: Ax\leq z\}
\end{equation}
for some $A\in\mathbb{R}^{m\times d}, z\in\mathbb{R}^m$.

The main theorem for polytopes \cite{Ziegler2012} states that the two descriptions are equivalent. The descriptions are also often called V-description and H-description, respectively. 

A face of $P$ is defined as 
\begin{equation*}
F \coloneqq \{x\in P: \alpha x\leq \alpha_0\}
\end{equation*}
with $\alpha$ and $\alpha_0$ real.

Then $F$ is facet defining if and only if
\begin{equation*}
\text{dim} F = \text{dim} P - 1
\end{equation*}

The minimal H-description of the polytope is its set of facets.

\subsection{Bell's local polytope}

To specify a Bell scenario, we fix the numbers of settings for Alice $x=1,\dots,X$, for Bob $y=1,\dots,Y$, and their numbers of outcomes $a=1,\dots,A$, $b=1,\dots,B$, respectively. Such a scenario is denoted $(X,Y,A,B)$.

The joint statistics arising from a Bell experiment are given by a probability distribution $p(ab|xy)$. Such a distribution is local when it satisfies \cite{Fine1982} :
\begin{equation}
    p(ab|xy) = \sum_{\lambda} p(\lambda) p^A(a|x\lambda)p^B(b|y\lambda)
\end{equation}
where $p^A(a|x\lambda), p^B(b|y\lambda) \in \{0,1\}$, i.e. they are deterministic strategies, labeled by $\lambda$, also called a local hidden variable. Given that these deterministic strategies are readily available to us (we just consider every combination of $p^A(a|x\lambda), p^B(b|y\lambda) \in \{0,1\}$) the complete V-description of Bell's local polytopes is easy to obtain.

\subsection{Collins-Gisin notation and classification of Bell inequalities}

In quantum theory, the joint statistics are obtained via the Born rule,
\begin{equation}
p(ab|xy) = \text{Tr}\Big(\Pi^A_{a|x}\otimes\Pi^B_{b|y}\rho\Big),
\end{equation}
where $\Pi^A_{a|x},\Pi^B_{b|y}$ are Alice and Bob's Positive Operator-Valued Measures (POVMs) and $\rho$ is a bipartite quantum state.

When we consider the set $\{p(ab|xy)\}_{abxy}$ as variables for the problem at-hand, for example facet enumeration, we say we use the P notation. The CG notation \cite{Collins2004} has the advantage of reducing the number of variables to the dimension of the polytope $(A-1)(B-1)XY + (A-1)X + (B-1)Y$ instead of considering a variable for every probability related to a scenario.  

Converting from the P to the CG notation is a two-step procedure. First, we use the no-signalling condition: add to your set of variables the following marginal distributions,
\begin{equation}
    p^A(a|x) \coloneqq \sum_{b=1}^B p(ab|xy^*), p^B(b|y) \coloneqq \sum_{a=1}^A p(ab|x^*y),
\end{equation}
where $x^*,y^*$ are arbitrary.

Secondly, we take advantage of the normalization (eq. \ref{eq:nrom}): remove all variables which include at least one outcome equal to the cardinality of that set of outcomes.

\begin{equation}
\label{eq:nrom}
    \sum_a p^A(a|x) = 1, \quad \sum_b p^B(b|y) = 1.
\end{equation}

A bipartite Bell inequality in CG notation thus takes the form,
\begin{align}
    \sum_{abxy}^{A-1,B-1}\alpha_{ab|xy}p(00|xy) & + \sum_{ax}^{A-1}\alpha^A_x p^A(a|x) \nonumber\\
    & \quad + \sum_{by}^{B-1}\alpha^B_y p^B(b|y) \leq L,
\end{align}
where when we do not add a superscript on the sum, it means we sum over all values the corresponding variable can take.

To classify the facets, we chose to convert them first to the P notation, where input and output symmetries correspond to simple permutations, and then convert them back to the CG notation. Any redundancy due to the normalization and the no-signalling conditions which could appear in the P notation is removed in the final step.

Finally the last symmetry that needs to be considered is when the scenario is completely symmetric ($X = Y$ and $A = B$), in which case we can switch Alice and Bob's labels. 

It is also important to select a representative amongst all facets that belong to a certain class. In this work, we used the lexicographic order, which speeds up facet enumeration in many cases \cite{Christof1996,Lorwald2015}. We arrange the CG notation elements in a single vector, where the first components are the joint probabilities, followed by Alice's marginals, and then Bob's marginals. The last element is the local bound. Hence, a given inequality is now expressed as,
\begin{equation}
    \Vec{\alpha}= [\alpha_{00|00}, \dots, \alpha^A_{0|0},\dots,\alpha^B_{0|0},\dots,L].
\end{equation}

Lexicographic (descending) order amounts to choosing as representative for the class the facet that satisfies the following relation:
\begin{equation}
\begin{split}
        \Vec{\alpha}^{'} >_\text{lex} \Vec{\alpha}^{''} \leftrightarrow \exists\ j \quad \text{s.t.}  \quad \forall \quad i < j, \\
    \Vec{\alpha}^{'}_i = \Vec{\alpha}^{''}_i, \quad
    \Vec{\alpha}^{'}_j > \Vec{\alpha}^{''}_j.
\end{split}
\end{equation}

\subsection{Slicing polytopes and facet enumeration}

For polytopes where direct facet enumeration is too difficult, we take inspiration from a method which has recently been used in \cite{Pal2010,Zambrini2019}, and extend it in a way that results in the discovery of many new tight Bell inequalities. The original procedure is simple: pick any hyperplane and adapt its bound such that it intersects the polytope. This will separate the set of vertices into two: the set of vertices which violate or saturate the hyperplane, and the set of vertices which don't. By modifying the bound, one can change the size of the resulting subpolytopes. Specifically, given the set of deterministic vertices $\{ \mathbf{x_k} \}$ that define the original polytope, we use the hyperplane $A$ such that $A$ divides the set $\{ \mathbf{x_k} \}$ into two sets, one formed by the deterministic vertices, $\{ \mathbf{x_k}^{'} \}$, which satisfy:
\begin{equation}
    A x_k^{'} \geq c
\end{equation}

\noindent and one composed of those that don't.

This results in a drastic decrease of the number of vertices, which lowers the complexity of the facet enumeration problem. One then simply solves a sufficiently small subpolytope, removes the artificial faces which are created by the slicing procedure, by verifying which separate the vertices of the original polytope and which have dimension $(A-1)(B-1)XY + (A-1)X + (B-1)Y-1$, to obtain a list of facets for the original scenario. 

\begin{figure}[H]
    \centering
    \includegraphics{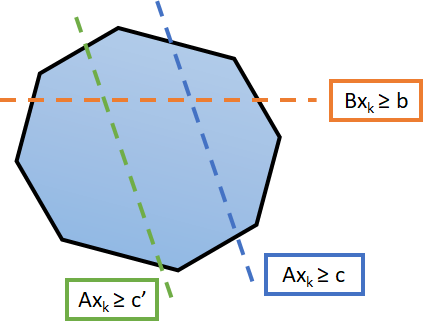}
    \caption{Representation of the polytope slices. By selecting different hyperplanes and bounds we can create subpolytopes with subsets of vertices, each containing some of the facets of the original polytope. }
    \label{fig:my_label}
\end{figure}

However, we find that frequently, simply using one slice does not allow to find the complete list of facets in a given scenario. Therefore, we extend our method to implement a series of slices, based on liftings \cite{pironio2005lifting} of already-known facets. With this systematic slicing procedure, we obtain a plethora of new tight Bell inequalities. Other extensions of the original slicing procedure could be considered, and it remains an open question what is the most efficient extension. In principle any hyperplane could be used. The question naturally arises which hyperplanes give cuts with more facets of the original polytope. Here, we show that using liftings of tight Bell inequalities in smaller scenarios is efficient. 

\par Given the high degree of complexity of the (4,4,2,2) scenario, which has 175 classes of facets for a total of 36 391 264 facets, and that the complete list of facets was already known, this poses an excellent benchmark test for our approach. Indeed, using a single inequality from the $I_{3322}$ class to provide a slice, we obtain 40,57\% (72 classes) of all the available classes. This sliced polytope was solved in less than 20 seconds, which clearly shows the strength of this approach, given that this computation would usually take several days. This allows for a more complete description, using 200 slices we found 75,43\% (132 classes).

For the facet enumeration, we use PANDA \cite{Lorwald2015}. This program will solve the facet enumeration problem faster when provided with a known list of facets. Hence, we use the facets we obtain via slicing to help us solve the facet enumeration problem for the original polytope.

\section{Properties of Interest} \label{sec:three}

\subsection{Quantum bounds}
The quantum bound $Q$ corresponds to the maximum value attainable by the inequality when using quantum resources. To compute this property two strategies were employed. Firstly, we use the NPA (Navascués, Pironio, and Acín) hierarchy \cite{NPAOrigin} (level 2) to compute a dimension agnostic upper bound to the maximum violation. The Ncpol2sdpa python package and the MOSEK solver \cite{mosek} are used to implement a hierarchy of Semi-Definite Programming (SDP) problems \cite{vandenberghe1996semidefinite}. 

Secondly, we use the python package CVXPY \cite{diamond2016cvxpy} to implement a seesaw of SDPs, generating a lower bound for Q. This allow us to generate dimension dependent quantum bounds and check how they compare against the value obtained using the NPA hierachy. In this work, we compute (lower) quantum bounds for qubits, $d=2$, qutrits, $d=3$ and ququarts, $d=4$.

\subsection{Resistance to noise}
Let $|\psi \rangle$ be a state which violates some Bell inequality maximally. The resistance to noise $\lambda$ is then defined as the minimum amount of white noise that needs to be mixed to the state $\rho$ such that the inequality no longer surpasses the local bound:

\begin{equation}
    |\rho \rangle = \lambda|\psi \rangle \langle \psi | + (1-\lambda) \frac{\mathbb{I}_{d^2}}{d^2}
\end{equation}

This quantity was calculated for qubits, $d=2$, qutrits, $d=3$ and ququarts, $d=4$.
\subsection{Detection efficiency}

When a detector fails  to trigger, Alice and Bob can choose which result to output. Hence there are $A^XB^Y$ strategies that the players may implement when their detectors fail. Consider the so-called symmetric scenario, where Bob and Alice have detectors with same detection efficiency $\eta$, we may now rewrite an inequality to reflect detector failure, following the approach of \cite{brunner2007detection},
\begin{equation}
    I(\eta) = \eta^2 Q +\eta(1-\eta)(M_A+M_B) +(1-\eta)^2Z
\end{equation}

\noindent where $Q$ represents the value obtained when both detectors work, $M_A$ represents the value obtained when Bob's detectors fail to activate, $M_B$ represents the value obtained when Alice's detectors fail to activate and $Z$ represents the value obtained when both detectors fail to trigger. The minimum detection efficiency were computed for the state and measurements that maximally violate the inequality.

By checking all the strategies, it is possible to extract the minimum detection efficiency for a given behavior, $\eta_{min}$, required to obtain a value above the local bound, $I(\eta_{min}) = L$, and close the detection loophole.

Once again this property is computed for all the dimensions considered.
\subsection{Concurrence}
For qubits, concurrence is a measure of entanglement of a state $\rho$ \cite{hill1997entanglement}, and can be defined as:
\begin{equation}
    C(\rho) = \max(0, \alpha_1 - \alpha_2 - \alpha_3 - \alpha_4),
\end{equation}

\noindent where $\alpha_i$ are the eigenvalues, in decreasing order, of the matrix $R$, defined as:
\begin{equation}
    R(\rho) = \sqrt{\sqrt{\rho} \Tilde{\rho} \sqrt{\rho}},
\end{equation}

\noindent where:

\begin{equation}
    \Tilde{\rho} = (\sigma_y \otimes \sigma_y) \rho^* (\sigma_y \otimes \sigma_y), 
\end{equation}

\noindent and $*$ denotes the complex conjugate. This property allows us to quantify how entangled the state that produces the maximum violation is.

\section{Results} \label{sec:four}

\subsection{(4,4,2,2)}

We revisit the 4422 scenario, which cannot be solved in a reasonable time (less than weeks) by direct facet enumeration\footnote{ One of us tried running facet enumeration for several weeks, no facets were produced.}. Using 200 cuts we first obtain a list of 132 classes of facets. Therefore, the only task left for PANDA is to find the remaining facets and prove that the list is complete. By giving PANDA the preliminary list, and symmetry information, PANDA is able to show the list is complete in approximately one day.

The properties of the classes of facets can be found in tables \ref{grid_A4422_2}, \ref{grid_A4422_3} and \ref{grid_A4422_4}.

\subsection{(6,3,2,2)}

In this scenario, there are no new facet classes, with respect to (5,3,2,2). Hence, there are six non-trivial facet classes. The properties of the classes of facets can be found in tables \ref{grid_A6322_2}, \ref{grid_A6322_3} and \ref{grid_A6322_4}.

Since the facet list is the same for $(5,3,2,2)$ and $(6,3,2,2)$, it seems plausible that for $(N,3,2,2)$ scenarios, where $N \geq 5$, are fully characterized by the six non-trivial facets mentioned above. We therefore conjecture that for arbitrary $N \geq 5$, $(N,3,2,2)$ is fully described by the six non trivial classes of facets. 

\subsection{(3,3,3,2)}

In this scenario, we find 25 classes, of which only 1 is non-lifted for a total of 252 558 facets. This list of classes was already presented in \cite{Cope2019}, but the authors conjectured that the list is complete, i.e. that it is the complete minimal H-description of the local set in this scenario. Our result proves that the conjecture is indeed correct. The properties of the classes of facets can be found in tables \ref{grid_A3332_2}, \ref{grid_A3332_3} and \ref{grid_A3332_4}. This result was achieved starting with a partial description obtained by using 200 slices, that provided 76 \% of the full description,

\subsection{(3,2,3,3)}
In this scenario we find 38 classes, of which only 5 are non-lifted for a total of 793 854 facets.
The properties of the classes of facets can be found in tables \ref{grid_A3233_2}, \ref{grid_A3233_3} and \ref{grid_A3233_4}.
\par To the best knowledge of the authors, this is the first time that the facet enumeration problem for this polytope was solved in the literature. This result was achieved starting with a partial description obtained by using 200 slices, that provided 65,79 \% of the full description,

\subsection{(3,3,4,2)}

In this scenario we find 159 classes for a total of 23 973 264  facets. It is possible that this list is complete given that no other classes were found after extensive slicing, nevertheless we are not able to prove it.
The properties of the classes of facets can be found in tables \ref{grid_A3342_2}, \ref{grid_A3342_3} and \ref{grid_A3342_4}.

\subsection{(2,2,4,4)}

In this scenario we find 34 classes for a total of 11 665 992 facets. Of these, 10 are non-lifted. No other classes were find after extensive slicing. The properties of the classes of facets can be found in tables \ref{grid_A2244_2}, \ref{grid_A2244_3} and \ref{grid_A2244_4}.

\subsection{(2,2,3,5)}

In this scenario we find 15 classes for a total of 286 260 facets. These correspond to the classes found in (2,2,3,3), therefore we did not find any new class with respect to that scenario.

\subsection{(4,3,3,2)}

In this scenario we find 80 classes for a total of 9 960 696 facets. Of these, only 1 is non-lifted. No other classes were find after extensive slicing. The properties of the classes of facets can be found in tables \ref{grid_A4332_2}, \ref{grid_A4332_3} and \ref{grid_A4332_4}.

\section{Promising Inequalities for quantum communication}\label{sec:five}

In Appendix \ref{app:A}, we present the most promising facets in terms of resistance to noise, minimum detection efficiency, or both. These facets demonstrate an advantage in terms of these quantities when compared to the simplest case, the CHSH inequality. Only a few such examples were known previously in the literature \cite{Gonzales2021}, here we present a list of 21 candidates. All the relevant quantities are computed for quantum states of dimension up to four. 

For example, the 21 candidates could show an advantage, compared to CHSH, in terms of rates for device-independent quantum key distribution (DI-QKD) \cite{Gonzales2021, brown2021computing, brown2021device}. Hence, our list of facets corresponds to 21 DI-QKD protocols which, in theory, may achieve better rates than CHSH. Nevertheless, one should also take into account that in practice, it is harder to generate and detect a qu$d$it ($d>2$) than a qubit. For a genuine advantage, one should therefore further consider the imperfections of state preparation and detection for different Hilbert space dimensions.

\section{Conclusions} 

Famously, finding complete lists of Bell inequalities for a given scenario via direct facet enumeration is a hard problem \cite{Pitowsky1989}. In this article, we show that one can go beyond direct facet enumeration using a simple procedure: slicing the polytope into smaller pieces using liftings of known tight Bell inequalities. Furthermore, the method is general and can readily be applied to any convex polytope.

We have used this simple procedure to find facets for Bell scenarios where direct facet enumeration does not work. We have derived the complete lists of facets for scenarios (6,3,2,2), (3,3,3,2), (3,2,3,3), and (2,2,3,5). In addition, we provide extensive lists of one representative per class of Bell inequalities in (2,2,4,4), (3,3,4,2) and (4,3,3,2). For all the facets presented in this work we also provide quantum bounds, resistance to noise, and detection efficiencies, for two-qudit systems with $d=2,3,4$. 

Fundamentally, our results help elucidate the general structure of Bell polytopes, hence understand more about Bell nonlocality in bipartite scenarios. For example, one could study the facets presented here to derive families of facets valid for larger numbers of settings and outcomes. Studying the geometry of the local set is indeed important to understand the limits of quantum nonlocality \cite{Goh2018,Le2021}, which is necessary both to answer fundamental questions, and to guarantee the security of quantum cryptographic protocols. For example, in what concerns DI-QKD, our results are promising in the perspective of inspiring new, improved, protocols.

\begin{acknowledgments}
 E.Z.C. acknowledges funding by FCT/MCTES - Funda\c c\~ao para a Ci\^encia e a Tecnologia (Portugal) - through national funds and when applicable co-funding by EU funds under the project UIDB/50008/2020.
\end{acknowledgments}

\bibliography{bibl}

\clearpage

\appendix

\section{Most promising facets}\label{app:A}

\onecolumngrid

{

\setlength{\LTcapwidth}{\dimexpr\linewidth+4.5in\relax}

\begin{center}\setlength\tabcolsep{16pt} 

\begin{longtable}{ c c c c c c c c }

\caption[Properties of the Facets of the most promising inequalities using qubits ($d=2$).]{Properties of the Facets of the most promising inequalities using qubits($d=2$).} \label{grid_A2} \\
\hline \multicolumn{1}{c}{Name} & \multicolumn{1}{c}{Scenario} & \multicolumn{1}{c}{$L$}& \multicolumn{1}{c}{$Q$}& \multicolumn{1}{c}{$Q_{NPA}$} & \multicolumn{1}{c}{$\lambda$} & \multicolumn{1}{c}{$\eta_{min}$} & \multicolumn{1}{c}{C}  \\ \hline 
\endfirsthead

\multicolumn{8}{c}%
{{\bfseries \tablename\ \thetable{} -- continued from previous page}} \\
\hline \multicolumn{1}{c}{Nº} & \multicolumn{1}{c}{Name} & \multicolumn{1}{c}{$L$}& \multicolumn{1}{c}{$Q$}& \multicolumn{1}{c}{$Q_{NPA}$} & \multicolumn{1}{c}{$\lambda$} & \multicolumn{1}{c}{$\eta_{min}$} & \multicolumn{1}{c}{C}  \\ \hline 
\endhead

\hline \multicolumn{8}{|r|}{{Continued on next page}} \\ \hline
\endfoot

\hline \hline
\endlastfoot
\hline

\text{CHSH} & (2,2,2,2) & 0.0&0.2071&0.2071&0.7071&0.8284&1.0 \\
\hline
$F_{6}$&(4,4,2,2) &0.0&0.25&0.2879&0.8&0.8284&1.0\\
\hline 
$F_{57}$&(4,4,2,2)&1.0&1.5436&1.643&0.7863&0.8449&0.9903\\
\hline 
$F_{72}$&(4,4,2,2)&0.0&0.4345&0.4554&0.8011&0.832&0.999\\
\hline 
$F_{79}$&(4,4,2,2)&1.0&1.6056&1.6056&0.7676&0.8282&0.9992\\
\hline 
$F_{80}$&(4,4,2,2)&0.0&0.4353&0.4353&0.7751&0.8178&0.9994\\
\hline
$F_{146}$&(4,4,2,2)&1.0&1.5932&1.5933&0.7468&0.8261&0.9978\\
\hline 
$I_{2233}$ & (2,2,3,3)&0&0.2071&0.305&0.7071&0.8284&1.0 \\
\hline

\end{longtable}

\end{center}

}

{

\setlength{\LTcapwidth}{\dimexpr\linewidth+4.5in\relax}

\begin{center}\setlength\tabcolsep{16pt} 

\begin{longtable}{ c c c c c c c }

\caption[Properties of the Facets of the most promising inequalities using  qutrits ($d=3$).]{Properties of the Facets of the most promising inequalities using  qutrits ($d=3$).} \label{grid_A3} \\
\hline \multicolumn{1}{c}{Name} & \multicolumn{1}{c}{Scenario} & \multicolumn{1}{c}{$L$}& \multicolumn{1}{c}{$Q$}& \multicolumn{1}{c}{$Q_{NPA}$} & \multicolumn{1}{c}{$\lambda$} & \multicolumn{1}{c}{$\eta_{min}$}   \\ \hline 
\endfirsthead

\multicolumn{7}{c}%
{{\bfseries \tablename\ \thetable{} -- continued from previous page}} \\
\hline \multicolumn{1}{c}{Name} & \multicolumn{1}{c}{Scenario} & \multicolumn{1}{c}{$L$}& \multicolumn{1}{c}{$Q$}& \multicolumn{1}{c}{$Q_{NPA}$} & \multicolumn{1}{c}{$\lambda$} & \multicolumn{1}{c}{$\eta_{min}$}  \\ \hline 
\endhead

\hline \multicolumn{7}{|r|}{{Continued on next page}} \\ \hline
\endfoot

\hline \hline
\endlastfoot
\hline

\text{CHSH} & (2,2,2,2)   &0.0&0.2071&0.2071&0.7071&0.8284\\
\hline
$F_{6}$&(4,4,2,2)&0.0&0.2604&0.2879&0.8175&0.8565\\
\hline
$F_{57}$&(4,4,2,2)&1.0&1.643&1.643&0.7917&0.8021\\
\hline
$F_{72}$&(4,4,2,2)&0.0&0.4554&0.4554&0.7735&0.8246\\
\hline
$F_{79}$&(4,4,2,2)&1.0&1.6056&1.6056&0.7909&0.8282\\
\hline 
$F_{80}$&(4,4,2,2)&0.0&0.4353&0.4353&0.7748&0.8178\\
\hline 
$F_{146}$&(4,4,2,2)&1.0&1.5932&1.5933&0.7964&0.8261\\
\hline
$I_{2233}$ & (2,2,3,3) &0&0.305&0.305&0.6861&0.8138 \\
\hline
\end{longtable}

\end{center}

}

{

\setlength{\LTcapwidth}{\dimexpr\linewidth+4.5in\relax}

\begin{center}\setlength\tabcolsep{16pt} 

\begin{longtable}{ c c c c c c c }

\caption[Properties of the Facets of the most promising inequalities using  ququarts ($d=4$).]{Properties of the Facets of the most promising inequalities using  ququarts ($d=4$).} \label{grid_A4} \\
\hline \multicolumn{1}{c}{Name} & \multicolumn{1}{c}{Scenario} & \multicolumn{1}{c}{$L$}& \multicolumn{1}{c}{$Q$}& \multicolumn{1}{c}{$Q_{NPA}$} & \multicolumn{1}{c}{$\lambda$} & \multicolumn{1}{c}{$\eta_{min}$}   \\ \hline 
\endfirsthead

\multicolumn{7}{c}%
{{\bfseries \tablename\ \thetable{} -- continued from previous page}} \\
\hline \multicolumn{1}{c}{Name} & \multicolumn{1}{c}{Scenario} & \multicolumn{1}{c}{$L$}& \multicolumn{1}{c}{$Q$}& \multicolumn{1}{c}{$Q_{NPA}$} & \multicolumn{1}{c}{$\lambda$} & \multicolumn{1}{c}{$\eta_{min}$}  \\ \hline 
\endhead

\hline \multicolumn{7}{|r|}{{Continued on next page}} \\ \hline
\endfoot

\hline \hline
\endlastfoot
\hline

\text{CHSH} & (2,2,2,2)   &0.0&0.2071&0.2071&0.7071&0.8284\\
\hline
$F_{6}$&(4,4,2,2)&0.0&0.2878&0.2879&0.8128&0.8181\\
\hline
$F_{57}$&(4,4,2,2)&1.0&1.643&1.643&0.811&0.8021\\
\hline
$F_{72}$&(4,4,2,2)&0.0&0.4554&0.4554&0.8049&0.8246\\
\hline 
$F_{79}$&(4,4,2,2)&1.0&1.6056&1.6056&0.7682&0.8282\\
\hline 
$F_{80}$&(4,4,2,2)&0.0&0.4353&0.4353&0.7747&0.8179\\
\hline 
$F_{146}$&(4,4,2,2)&1.0&1.5932&1.5933&0.8018&0.8261\\
\hline 
$I_{2233}$&(2,2,3,3)&0&0.305&0.305&0.6861&0.8138\\
\hline
\end{longtable}

\end{center}

}

\clearpage

\section{Analysis of facet lists}\label{app:B}

\onecolumngrid

\subsection{(4,4,2,2)}

{

\setlength{\LTcapwidth}{\dimexpr\linewidth+4.5in\relax}

\begin{center}\setlength\tabcolsep{16pt} 

\begin{longtable}{ c c c c c c c c }

\caption[Properties of the Facets of the (4,4,2,2) scenario using qubits ($d=2$).]{Properties of the Facets of the (4,4,2,2) scenario using qubits ($d=2$).} \label{grid_A4422_2} \\
\hline \multicolumn{1}{c}{Nº} & \multicolumn{1}{c}{Name} & \multicolumn{1}{c}{$L$}& \multicolumn{1}{c}{$Q$}& \multicolumn{1}{c}{$Q_{NPA}$} & \multicolumn{1}{c}{$\lambda$} & \multicolumn{1}{c}{$\eta_{min}$} & \multicolumn{1}{c}{C}  \\ \hline 
\endfirsthead

\multicolumn{8}{c}%
{{\bfseries \tablename\ \thetable{} -- continued from previous page}} \\
\hline \multicolumn{1}{c}{Nº} & \multicolumn{1}{c}{Name} & \multicolumn{1}{c}{$L$}& \multicolumn{1}{c}{$Q$}& \multicolumn{1}{c}{$Q_{NPA}$} & \multicolumn{1}{c}{$\lambda$} & \multicolumn{1}{c}{$\eta_{min}$} & \multicolumn{1}{c}{C}  \\ \hline 
\endhead

\hline \multicolumn{8}{|r|}{{Continued on next page}} \\ \hline
\endfoot

\hline \hline
\endlastfoot
\hline

1 & \text{CHSH} & 0.0&0.2071&0.2071&0.7071&0.8284&1.0 \\
\hline
2 & $I_{3322}$ & 0.0&0.25&0.2509&0.8&0.8284&1.0 \\
\hline
3 & $I_{4322}^1$ & 0.0&0.4142&0.4142&0.7071&0.8284&1.0 \\
\hline
4 & $I_{4322}^2$ & 1.0&1.299&1.29904&0.7698&0.8699&1.0 \\
\hline
5 & $I_{4322}^3$ & 1.0&1.4365&1.4365&0.7746&0.8513&1.0 \\
\hline
6&$F_{1}$&1.0&1.8785&1.8785&0.74&0.8506&1.0\\
\hline 
7&$F_{2}$&1.0&1.5412&1.5412&0.7349&0.8472&1.0\\
\hline 
8&$F_{3}$&0.0&0.4142&0.4495&0.8284&0.8787&1.0\\
\hline

\end{longtable}

\end{center}

}

{

\setlength{\LTcapwidth}{\dimexpr\linewidth+4.5in\relax}

\begin{center}\setlength\tabcolsep{16pt} 

\begin{longtable}{ c c c c c c c }

\caption[Properties of the Facets of the (4,4,2,2) scenario using qutrits ($d=3$).]{Properties of the Facets of the (4,4,2,2) scenario using qutrits ($d=3$).} \label{grid_A4422_3} \\
\hline \multicolumn{1}{c}{Nº} & \multicolumn{1}{c}{Name} & \multicolumn{1}{c}{$L$}& \multicolumn{1}{c}{$Q$}& \multicolumn{1}{c}{$Q_{NPA}$} & \multicolumn{1}{c}{$\lambda$} & \multicolumn{1}{c}{$\eta_{min}$}   \\ \hline 
\endfirsthead

\multicolumn{7}{c}%
{{\bfseries \tablename\ \thetable{} -- continued from previous page}} \\
\hline \multicolumn{1}{c}{Nº} & \multicolumn{1}{c}{Name} & \multicolumn{1}{c}{$L$}& \multicolumn{1}{c}{$Q$}& \multicolumn{1}{c}{$Q_{NPA}$} & \multicolumn{1}{c}{$\lambda$} & \multicolumn{1}{c}{$\eta_{min}$}  \\ \hline 
\endhead

\hline \multicolumn{7}{|r|}{{Continued on next page}} \\ \hline
\endfoot

\hline \hline
\endlastfoot
\hline

1 & \text{CHSH} &0.0&0.2071&0.2071&0.7071&0.8284 \\
\hline
2 & $I_{3322}$ & 0.0&0.25&0.2509&0.7995&0.8284 \\
\hline
3 & $I_{4322}^1$ & 0.0&0.4142&0.4142&0.745&0.8284 \\
\hline
4 & $I_{4322}^2$ & 1.0&1.299&1.2990&0.7878&0.8699 \\
\hline
5 & $I_{4322}^3$ & 1.0&1.4365&1.4365&0.7745&0.8513 \\
\hline
6&$F_{1}$&1.0&1.8785&1.8785&0.74&0.8506\\
\hline 
7&$F_{2}$&1.0&1.5412&1.5412&0.733&0.8472\\
\hline 
8&$F_{3}$&0.0&0.4142&0.4495&0.8284&0.8787\\
\hline 

\end{longtable}

\end{center}

}

{

\setlength{\LTcapwidth}{\dimexpr\linewidth+4.5in\relax}

\begin{center}\setlength\tabcolsep{16pt} 

\begin{longtable}{ c c c c c c c }

\caption[Properties of the Facets of the (4,4,2,2) scenario using ququarts ($d=4$).]{Properties of the Facets of the (4,4,2,2) scenario using ququarts ($d=4$).} \label{grid_A4422_4} \\
\hline \multicolumn{1}{c}{Nº} & \multicolumn{1}{c}{Name} & \multicolumn{1}{c}{$L$}& \multicolumn{1}{c}{$Q$}& \multicolumn{1}{c}{$Q_{NPA}$} & \multicolumn{1}{c}{$\lambda$} & \multicolumn{1}{c}{$\eta_{min}$}  \\ \hline 
\endfirsthead

\multicolumn{7}{c}%
{{\bfseries \tablename\ \thetable{} -- continued from previous page} }\\
\hline \multicolumn{1}{c}{Nº} & \multicolumn{1}{c}{Name} & \multicolumn{1}{c}{$L$}& \multicolumn{1}{c}{$Q$}& \multicolumn{1}{c}{$Q_{NPA}$} & \multicolumn{1}{c}{$\lambda$} & \multicolumn{1}{c}{$\eta_{min}$}   \\ \hline 
\endhead

\hline \multicolumn{7}{|r|}{{Continued on next page}} \\ \hline
\endfoot

\hline \hline
\endlastfoot
\hline

1 & \text{CHSH} & 0&0.2071&0.2071&0.7071&0.8284 \\
\hline
2 & $I_{3322}$ & 0&0.25&0.2509&0.7994&0.8284 \\
\hline
3 & $I_{4322}^1$ & 0&0.4142&0.4142&0.7553&0.8284 \\
\hline
4 & $I_{4322}^2$ & 1&1.299&1.2990&0.7926&0.8699 \\
\hline
5 & $I_{4322}^3$ & 1&1.4365&1.43649&0.7746&0.8514 \\
\hline
6&$F_{1}$&1.0&1.8785&1.8785&0.74&0.8506\\
\hline 
7&$F_{2}$&1.0&1.5412&1.5412&0.7349&0.8472\\
\hline 
8&$F_{3}$&0.0&0.4142&0.4495&0.7898&0.8284\\
\hline

\end{longtable}
\end{center}
}

\subsection{(6,3,2,2)}

{

\setlength{\LTcapwidth}{\dimexpr\linewidth+4.5in\relax}

\begin{center}\setlength\tabcolsep{16pt} 

\begin{longtable}{ c c c c c c c c }

\caption[Properties of the Facets of the (6,3,2,2) scenario using qubits ($d=2$).]{Properties of the Facets of the (6,3,2,2) scenario using qubits ($d=2$).} \label{grid_A6322_2} \\
\hline \multicolumn{1}{c}{Nº} & \multicolumn{1}{c}{Name} & \multicolumn{1}{c}{$L$}& \multicolumn{1}{c}{$Q$}& \multicolumn{1}{c}{$Q_{NPA}$} & \multicolumn{1}{c}{$\lambda$} & \multicolumn{1}{c}{$\eta_{min}$} & \multicolumn{1}{c}{C}  \\ \hline 
\endfirsthead

\multicolumn{8}{c}%
{{\bfseries \tablename\ \thetable{} -- continued from previous page}} \\
\hline \multicolumn{1}{c}{Nº} & \multicolumn{1}{c}{Name} & \multicolumn{1}{c}{$L$}& \multicolumn{1}{c}{$Q$}& \multicolumn{1}{c}{$Q_{NPA}$} & \multicolumn{1}{c}{$\lambda$} & \multicolumn{1}{c}{$\eta_{min}$} & \multicolumn{1}{c}{C}  \\ \hline 
\endhead

\hline \multicolumn{8}{|r|}{{Continued on next page}} \\ \hline
\endfoot

\hline \hline
\endlastfoot
\hline

1 & \text{CHSH} & 0.0&0.2071&0.2071&0.7071&0.8284&1.0 \\
\hline
2 & $I_{3322}$ & 0.0&0.25&0.2509&0.8&0.8284&1.0 \\
\hline
3 & $I_{4322}^1$ & 0.0&0.4142&0.4142&0.7071&0.8284&1.0 \\
\hline
4 & $I_{4322}^2$ & 1.0&1.299&1.29904&0.7698&0.8699&1.0 \\
\hline
5 & $I_{4322}^3$ & 1.0&1.4365&1.4365&0.7746&0.8513&1.0 \\
\hline
6 & $I_{5322}$& 0&0.4485&0.4843&0.7698&0.8479&1.0 \\
\hline
\end{longtable}
\end{center}
}

{

\setlength{\LTcapwidth}{\dimexpr\linewidth+4.5in\relax}

\begin{center}\setlength\tabcolsep{16pt} 

\begin{longtable}{ c c c c c c c }

\caption[Properties of the Facets of the (6,3,2,2) scenario using qutrits ($d=3$).]{Properties of the Facets of the (6,3,2,2) scenario using qutrits ($d=3$).} \label{grid_A6322_3} \\
\hline \multicolumn{1}{c}{Nº} & \multicolumn{1}{c}{Name} & \multicolumn{1}{c}{$L$}& \multicolumn{1}{c}{$Q$}& \multicolumn{1}{c}{$Q_{NPA}$} & \multicolumn{1}{c}{$\lambda$} & \multicolumn{1}{c}{$\eta_{min}$}   \\ \hline 
\endfirsthead

\multicolumn{7}{c}%
{{\bfseries \tablename\ \thetable{} -- continued from previous page}} \\
\hline \multicolumn{1}{c}{Nº} & \multicolumn{1}{c}{Name} & \multicolumn{1}{c}{$L$}& \multicolumn{1}{c}{$Q$}& \multicolumn{1}{c}{$Q_{NPA}$} & \multicolumn{1}{c}{$\lambda$} & \multicolumn{1}{c}{$\eta_{min}$}  \\ \hline 
\endhead

\hline \multicolumn{7}{|r|}{{Continued on next page}} \\ \hline
\endfoot

\hline \hline
\endlastfoot
\hline

1 & \text{CHSH} &0.0&0.2071&0.2071&0.7071&0.8284 \\
\hline
2 & $I_{3322}$ & 0.0&0.25&0.2509&0.7995&0.8284 \\
\hline
3 & $I_{4322}^1$ & 0.0&0.4142&0.4142&0.745&0.8284 \\
\hline
4 & $I_{4322}^2$ & 1.0&1.299&1.2990&0.7878&0.8699 \\
\hline
5 & $I_{4322}^3$ & 1.0&1.4365&1.4365&0.7745&0.8513 \\
\hline
6 & $I_{5322}$& 0&0.4485&0.4843&0.7704&0.8479 \\
\hline
\end{longtable}
\end{center}
}

{

\setlength{\LTcapwidth}{\dimexpr\linewidth+4.5in\relax}

\begin{center}\setlength\tabcolsep{16pt} 

\begin{longtable}{ c c c c c c c }

\caption[Properties of the Facets of the (6,3,2,2) scenario using ququarts ($d=4$).]{Properties of the Facets of the (6,3,2,2) scenario using ququarts ($d=4$).} \label{grid_A6322_4} \\
\hline \multicolumn{1}{c}{Nº} & \multicolumn{1}{c}{Name} & \multicolumn{1}{c}{$L$}& \multicolumn{1}{c}{$Q$}& \multicolumn{1}{c}{$Q_{NPA}$} & \multicolumn{1}{c}{$\lambda$} & \multicolumn{1}{c}{$\eta_{min}$}  \\ \hline 
\endfirsthead

\multicolumn{7}{c}%
{{\bfseries \tablename\ \thetable{} -- continued from previous page} }\\
\hline \multicolumn{1}{c}{Nº} & \multicolumn{1}{c}{Name} & \multicolumn{1}{c}{$L$}& \multicolumn{1}{c}{$Q$}& \multicolumn{1}{c}{$Q_{NPA}$} & \multicolumn{1}{c}{$\lambda$} & \multicolumn{1}{c}{$\eta_{min}$}   \\ \hline 
\endhead

\hline \multicolumn{7}{|r|}{{Continued on next page}} \\ \hline
\endfoot

\hline \hline
\endlastfoot
\hline

1 & \text{CHSH} & 0&0.2071&0.2071&0.7071&0.8284 \\
\hline
2 & $I_{3322}$ & 0&0.25&0.2509&0.7994&0.8284 \\
\hline
3 & $I_{4322}^1$ & 0&0.4142&0.4142&0.7553&0.8284 \\
\hline
4 & $I_{4322}^2$ & 1&1.299&1.2990&0.7926&0.8699 \\
\hline
5 & $I_{4322}^3$ & 1&1.4365&1.43649&0.7746&0.8514 \\
\hline
6 & $I_{5322}$ & 0&0.4519&0.4843&0.7685&0.8469 \\
\hline

\end{longtable}
\end{center}
}

\subsection{(3,3,3,2)}

{

\setlength{\LTcapwidth}{\dimexpr\linewidth+4.5in\relax}

\begin{center}\setlength\tabcolsep{16pt} 

\begin{longtable}{ c c c c c c c c }

\caption[Properties of the Facets of the (3,3,3,2) scenario using qubits ($d=2$).]{Properties of the Facets of the (3,3,3,2) scenario using qubits ($d=2$).} \label{grid_A3332_2} \\
\hline \multicolumn{1}{c}{Nº} & \multicolumn{1}{c}{Name} & \multicolumn{1}{c}{$L$}& \multicolumn{1}{c}{$Q$}& \multicolumn{1}{c}{$Q_{NPA}$} & \multicolumn{1}{c}{$\lambda$} & \multicolumn{1}{c}{$\eta_{min}$} & \multicolumn{1}{c}{C}  \\ \hline 
\endfirsthead

\multicolumn{8}{c}%
{{\bfseries \tablename\ \thetable{} -- continued from previous page}} \\
\hline \multicolumn{1}{c}{Nº} & \multicolumn{1}{c}{Name} & \multicolumn{1}{c}{$L$}& \multicolumn{1}{c}{$Q$}& \multicolumn{1}{c}{$Q_{NPA}$} & \multicolumn{1}{c}{$\lambda$} & \multicolumn{1}{c}{$\eta_{min}$} & \multicolumn{1}{c}{C}  \\ \hline 
\endhead

\hline \multicolumn{8}{|r|}{{Continued on next page}} \\ \hline
\endfoot

\hline \hline
\endlastfoot
\hline
1&\text{CHSH}   &0.0&0.2071&0.2071&0.7071&0.8284&1.0\\
\hline 
2&$I_{[[2,3],[2,2,2]]}$   &0.0&0.2071&0.2532&0.7635&0.8284&1.0\\
\hline 
3&$I_{[[2,3],[2,2,2]]}$    &0.0&0.2071&0.2532&0.7635&0.8284&1.0\\
\hline 
4&$I_{2332}$ &0.0&0.4142&0.4142&0.7071&0.8284&1.0\\
\hline 
5&$I_{3322}$   &0.0&0.25&0.2509&0.8&0.8284&1.0\\
\hline 
6&$I_{3322}$   &0.0&0.25&0.2509&0.8&0.8284&1.0\\
\hline 
7&$I_{3322}$   &0.0&0.25&0.2509&0.8&0.8284&1.0\\
\hline 
8&$I_{[[2,2,3],[2,2,2]]}^1$   &0.0&0.4142&0.4142&0.7071&0.8284&1.0\\
\hline 

\end{longtable}

\end{center}

}

{

\setlength{\LTcapwidth}{\dimexpr\linewidth+4.5in\relax}

\begin{center}\setlength\tabcolsep{16pt} 

\begin{longtable}{ c c c c c c c }

\caption[Properties of the Facets of the (3,3,3,2) scenario using qutrits ($d=3$).]{Properties of the Facets of the (3,3,3,2) scenario using qutrits ($d=3$).} \label{grid_A3332_3} \\
\hline \multicolumn{1}{c}{Nº} & \multicolumn{1}{c}{Name} & \multicolumn{1}{c}{$L$}& \multicolumn{1}{c}{$Q$}& \multicolumn{1}{c}{$Q_{NPA}$} & \multicolumn{1}{c}{$\lambda$} & \multicolumn{1}{c}{$\eta_{min}$}   \\ \hline 
\endfirsthead

\multicolumn{7}{c}%
{{\bfseries \tablename\ \thetable{} -- continued from previous page}} \\
\hline \multicolumn{1}{c}{Nº} & \multicolumn{1}{c}{Name} & \multicolumn{1}{c}{$L$}& \multicolumn{1}{c}{$Q$}& \multicolumn{1}{c}{$Q_{NPA}$} & \multicolumn{1}{c}{$\lambda$} & \multicolumn{1}{c}{$\eta_{min}$}  \\ \hline 
\endhead

\hline \multicolumn{7}{|r|}{{Continued on next page}} \\ \hline
\endfoot

\hline \hline
\endlastfoot
\hline

1&\text{CHSH}   &0.0&0.2071&0.2071&0.7071&0.8284\\
\hline 
2&$I_{[[2,3],[2,2,2]]}$   &0.0&0.2532&0.2532&0.7247&0.8301\\
\hline 
3&$I_{[[2,3],[2,2,2]]}$    &0.0&0.2532&0.2532&0.7247&0.83\\
\hline 
4&$I_{2332}$ &0.0&0.4142&0.4142&0.7288&0.8284\\
\hline 
5&$I_{3322}$   &0.0&0.25&0.2509&0.8057&0.8284\\
\hline 
6&$I_{3322}$   &0.0&0.25&0.2509&0.8096&0.8284\\
\hline 
7&$I_{3322}$   &0.0&0.25&0.2509&0.8351&0.8284\\
\hline 
8&$I_{[[2,2,3],[2,2,2]]}^1$   &0.0&0.4142&0.4142&0.7333&0.8284\\
\hline 

\end{longtable}

\end{center}

}

{

\setlength{\LTcapwidth}{\dimexpr\linewidth+4.5in\relax}

\begin{center}\setlength\tabcolsep{16pt} 

\begin{longtable}{ c c c c c c c }

\caption[Properties of the Facets of the (3,3,3,2) scenario using ququarts ($d=4$).]{Properties of the Facets of the (3,3,3,2) scenario using ququarts ($d=4$).} \label{grid_A3332_4} \\
\hline \multicolumn{1}{c}{Nº} & \multicolumn{1}{c}{Name} & \multicolumn{1}{c}{$L$}& \multicolumn{1}{c}{$Q$}& \multicolumn{1}{c}{$Q_{NPA}$} & \multicolumn{1}{c}{$\lambda$} & \multicolumn{1}{c}{$\eta_{min}$}   \\ \hline 
\endfirsthead

\multicolumn{7}{c}%
{{\bfseries \tablename\ \thetable{} -- continued from previous page}} \\
\hline \multicolumn{1}{c}{Nº} & \multicolumn{1}{c}{Name} & \multicolumn{1}{c}{$L$}& \multicolumn{1}{c}{$Q$}& \multicolumn{1}{c}{$Q_{NPA}$} & \multicolumn{1}{c}{$\lambda$} & \multicolumn{1}{c}{$\eta_{min}$}  \\ \hline 
\endhead

\hline \multicolumn{7}{|r|}{{Continued on next page}} \\ \hline
\endfoot

\hline \hline
\endlastfoot
\hline
1&\text{CHSH}   &0.0&0.2071&0.2071&0.7071&0.8284\\
\hline 
2&$I_{[[2,3],[2,2,2]]}$   &0.0&0.2532&0.2532&0.735&0.8301\\
\hline 
3&$I_{[[2,3],[2,2,2]]}$    &0.0&0.2532&0.2532&0.7389&0.8301\\
\hline 
4&$I_{2332}$ &0.0&0.4142&0.4142&0.7387&0.8284\\
\hline 
5&$I_{3322}$   &0.0&0.25&0.2509&0.783&0.8284\\
\hline 
6&$I_{3322}$   &0.0&0.25&0.2509&0.8129&0.8284\\
\hline 
7&$I_{3322}$   &0.0&0.25&0.2509&0.8283&0.8284\\
\hline 
8&$I_{[[2,2,3],[2,2,2]]}^1$   &0.0&0.4142&0.4142&0.739&0.8284\\
\hline 

\end{longtable}
\end{center}
}

\subsection{(3,2,3,3)}

{

\setlength{\LTcapwidth}{\dimexpr\linewidth+4.5in\relax}

\begin{center}\setlength\tabcolsep{16pt} 

\begin{longtable}{ c c c c c c c c }

\caption[Properties of the Facets of the (3,2,3,3) scenario using qubits ($d=2$).]{Properties of the Facets of the (3,2,3,3) scenario using qubits ($d=2$).} \label{grid_A3233_2} \\
\hline \multicolumn{1}{c}{Nº} & \multicolumn{1}{c}{Name} & \multicolumn{1}{c}{$L$}& \multicolumn{1}{c}{$Q$}& \multicolumn{1}{c}{$Q_{NPA}$} & \multicolumn{1}{c}{$\lambda$} & \multicolumn{1}{c}{$\eta_{min}$} & \multicolumn{1}{c}{C}  \\ \hline 
\endfirsthead

\multicolumn{8}{c}%
{{\bfseries \tablename\ \thetable{} -- continued from previous page}} \\
\hline \multicolumn{1}{c}{Nº} & \multicolumn{1}{c}{Name} & \multicolumn{1}{c}{$L$}& \multicolumn{1}{c}{$Q$}& \multicolumn{1}{c}{$Q_{NPA}$} & \multicolumn{1}{c}{$\lambda$} & \multicolumn{1}{c}{$\eta_{min}$} & \multicolumn{1}{c}{C}  \\ \hline 
\endhead

\hline \multicolumn{8}{|r|}{{Continued on next page}} \\ \hline
\endfoot

\hline \hline
\endlastfoot
\hline
1&\text{CHSH}   &0.0&0.2071&0.2071&0.7071&0.8284&1.0\\
\hline 
2&\text{CHSH}   &0.0&0.2071&0.2071&0.7071&0.8284&1.0\\
\hline 
3&$I_{[[2,2,2],[2,3]]}$   &0.0&0.2071&0.2532&0.7488&0.8284&1.0\\
\hline 
4&$I_{[[2,2,2],[2,3]]}$   &0.0&0.2071&0.2532&0.7725&0.8284&1.0\\
\hline 
5&$I_{[[2,2,2],[2,3]]}$   &0.0&0.2071&0.2532&0.7489&0.8284&1.0\\
\hline 
6&$I_{[[2,2,2],[2,3]]}$   &0.0&0.2071&0.2532&0.7725&0.8283&1.0\\
\hline 
7&$I_{[[2,2,2],[2,3]]}$   &0.0&0.2071&0.2532&0.7488&0.8284&1.0\\
\hline 
8&$I_{[[2,2,2],[2,3]]}$   &0.0&0.2071&0.2532&0.7725&0.8284&1.0\\
\hline 

\end{longtable}

\end{center}
}

{

\setlength{\LTcapwidth}{\dimexpr\linewidth+4.5in\relax}

\begin{center}\setlength\tabcolsep{16pt} 

\begin{longtable}{ c c c c c c c  }

\caption[Properties of the Facets of the (3,2,3,3) scenario using qutrits ($d=3$).]{Properties of the Facets of the (3,2,3,3) scenario using qutrits ($d=3$).} \label{grid_A3233_3} \\
\hline \multicolumn{1}{c}{Nº} & \multicolumn{1}{c}{Name} & \multicolumn{1}{c}{$L$}& \multicolumn{1}{c}{$Q$}& \multicolumn{1}{c}{$Q_{NPA}$} & \multicolumn{1}{c}{$\lambda$} & \multicolumn{1}{c}{$\eta_{min}$}   \\ \hline 
\endfirsthead

\multicolumn{7}{c}%
{{\bfseries \tablename\ \thetable{} -- continued from previous page}} \\
\hline \multicolumn{1}{c}{Nº} & \multicolumn{1}{c}{Name} & \multicolumn{1}{c}{$L$}& \multicolumn{1}{c}{$Q$}& \multicolumn{1}{c}{$Q_{NPA}$} & \multicolumn{1}{c}{$\lambda$} & \multicolumn{1}{c}{$\eta_{min}$}  \\ \hline 
\endhead

\hline \multicolumn{7}{|r|}{{Continued on next page}} \\ \hline
\endfoot

\hline \hline
\endlastfoot
\hline

1&\text{CHSH}   &0.0&0.2071&0.2071&0.7103&0.8284\\
\hline 
2&\text{CHSH}   &0.0&0.2071&0.2071&0.7039&0.8284\\
\hline 
3&$I_{[[2,2,2],[2,3]]}$   &0.0&0.2532&0.2532&0.7247&0.8301\\
\hline 
4&$I_{[[2,2,2],[2,3]]}$   &0.0&0.2532&0.2532&0.7247&0.8301\\
\hline 
5&$I_{[[2,2,2],[2,3]]}$   &0.0&0.2532&0.2532&0.7247&0.8301\\
\hline 
6&$I_{[[2,2,2],[2,3]]}$   &0.0&0.2532&0.2532&0.7247&0.8301\\
\hline 
7&$I_{[[2,2,2],[2,3]]}$   &0.0&0.2532&0.2532&0.7247&0.8301\\
\hline 
8&$I_{[[2,2,2],[2,3]]}$   &0.0&0.2532&0.2532&0.7247&0.8301\\
\hline

\end{longtable}

\end{center}
}

{

\setlength{\LTcapwidth}{\dimexpr\linewidth+4.5in\relax}

\begin{center}\setlength\tabcolsep{16pt} 

\begin{longtable}{ c c c c c c c  }

\caption[Properties of the Facets of the (3,2,3,3) scenario using ququarts ($d=4$).]{Properties of the Facets of the (3,2,3,3) scenario using ququarts ($d=4$).} \label{grid_A3233_4} \\
\hline \multicolumn{1}{c}{Nº} & \multicolumn{1}{c}{Name} & \multicolumn{1}{c}{$L$}& \multicolumn{1}{c}{$Q$}& \multicolumn{1}{c}{$Q_{NPA}$} & \multicolumn{1}{c}{$\lambda$} & \multicolumn{1}{c}{$\eta_{min}$}  \\ \hline 
\endfirsthead

\multicolumn{7}{c}%
{{\bfseries \tablename\ \thetable{} -- continued from previous page}} \\
\hline \multicolumn{1}{c}{Nº} & \multicolumn{1}{c}{Name} & \multicolumn{1}{c}{$L$}& \multicolumn{1}{c}{$Q$}& \multicolumn{1}{c}{$Q_{NPA}$} & \multicolumn{1}{c}{$\lambda$} & \multicolumn{1}{c}{$\eta_{min}$}   \\ \hline 
\endhead

\hline \multicolumn{7}{|r|}{{Continued on next page}} \\ \hline
\endfoot

\hline \hline
\endlastfoot
\hline

1&\text{CHSH}   &0.0&0.2071&0.2071&0.7142&0.8284\\
\hline 
2&\text{CHSH}   &0.0&0.2071&0.2071&0.7232&0.8284\\
\hline 
3&$I_{[[2,2,2],[2,3]]}$   &0.0&0.2532&0.2532&0.7245&0.8301\\
\hline 
4&$I_{[[2,2,2],[2,3]]}$   &0.0&0.2532&0.2532&0.7512&0.8301\\
\hline 
5&$I_{[[2,2,2],[2,3]]}$   &0.0&0.2532&0.2532&0.7245&0.8301\\
\hline 
6&$I_{[[2,2,2],[2,3]]}$   &0.0&0.2532&0.2532&0.744&0.8301\\
\hline 
7&$I_{[[2,2,2],[2,3]]}$   &0.0&0.2532&0.2532&0.7313&0.8301\\
\hline 
8&$I_{[[2,2,2],[2,3]]}$   &0.0&0.2532&0.2532&0.7378&0.8301\\
\hline 
\end{longtable}

\end{center}
}

\subsection{(3,3,4,2)}
{

\setlength{\LTcapwidth}{\dimexpr\linewidth+4.5in\relax}

\begin{center}\setlength\tabcolsep{16pt} 

\begin{longtable}{ c c c c c c c c }

\caption[Properties of the Facets of the (3,3,4,2) scenario using qubits ($d=2$).]{Properties of the Facets of the (3,3,4,2) scenario using qubits ($d=2$).} \label{grid_A3342_2} \\
\hline \multicolumn{1}{c}{Nº} & \multicolumn{1}{c}{Name} & \multicolumn{1}{c}{$L$}& \multicolumn{1}{c}{$Q$}& \multicolumn{1}{c}{$Q_{NPA}$} & \multicolumn{1}{c}{$\lambda$} & \multicolumn{1}{c}{$\eta_{min}$} & \multicolumn{1}{c}{C}  \\ \hline 
\endfirsthead

\multicolumn{8}{c}%
{{\bfseries \tablename\ \thetable{} -- continued from previous page}} \\
\hline \multicolumn{1}{c}{Nº} & \multicolumn{1}{c}{Name} & \multicolumn{1}{c}{$L$}& \multicolumn{1}{c}{$Q$}& \multicolumn{1}{c}{$Q_{NPA}$} & \multicolumn{1}{c}{$\lambda$} & \multicolumn{1}{c}{$\eta_{min}$} & \multicolumn{1}{c}{C}  \\ \hline 
\endhead

\hline \multicolumn{8}{|r|}{{Continued on next page}} \\ \hline
\endfoot

\hline \hline
\endlastfoot
\hline
1&\text{CHSH}&0.0&0.2071&0.2071&0.7071&0.8284&1.0\\
\hline 
2&\text{CHSH}&0.0&0.2071&0.2071&0.7071&0.8284&1.0\\
\hline 
3&\text{CHSH}&0.0&0.2071&0.2071&0.7071&0.8284&1.0\\
\hline 
4&$I_{3322}$&0.0&0.25&0.2509&0.8&0.8284&1.0\\
\hline 
5&$I_{3322}$&0.0&0.25&0.2509&0.8&0.8284&1.0\\
\hline 
6&$I_{3322}$&0.0&0.25&0.2509&0.8&0.8284&1.0\\
\hline 
7&$I_{3322}$&0.0&0.25&0.2509&0.8&0.8284&1.0\\
\hline 
8&$I_{3322}$&0.0&0.25&0.2509&0.8&0.8284&1.0\\
\hline 

\end{longtable}

\end{center}
}

{

\setlength{\LTcapwidth}{\dimexpr\linewidth+4.5in\relax}

\begin{center}\setlength\tabcolsep{16pt} 

\begin{longtable}{ c c c c c c c  }

\caption[Properties of the Facets of the (3,3,4,2) scenario using qutrits ($d=3$).]{Properties of the Facets of the (3,3,4,2) scenario using qutrits ($d=3$).} \label{grid_A3342_3} \\
\hline \multicolumn{1}{c}{Nº} & \multicolumn{1}{c}{Name} & \multicolumn{1}{c}{$L$}& \multicolumn{1}{c}{$Q$}& \multicolumn{1}{c}{$Q_{NPA}$} & \multicolumn{1}{c}{$\lambda$} & \multicolumn{1}{c}{$\eta_{min}$}   \\ \hline 
\endfirsthead

\multicolumn{7}{c}%
{{\bfseries \tablename\ \thetable{} -- continued from previous page}} \\
\hline \multicolumn{1}{c}{Nº} & \multicolumn{1}{c}{Name} & \multicolumn{1}{c}{$L$}& \multicolumn{1}{c}{$Q$}& \multicolumn{1}{c}{$Q_{NPA}$} & \multicolumn{1}{c}{$\lambda$} & \multicolumn{1}{c}{$\eta_{min}$}  \\ \hline 
\endhead

\hline \multicolumn{7}{|r|}{{Continued on next page}} \\ \hline
\endfoot

\hline \hline
\endlastfoot
\hline
1&\text{CHSH}&0.0&0.2071&0.2071&0.7071&0.8284\\
\hline 
2&\text{CHSH}&0.0&0.2071&0.2071&0.7071&0.8284\\
\hline 
3&\text{CHSH}&0.0&0.2071&0.2071&0.7071&0.8284\\
\hline 
4&$I_{3322}$&0.0&0.25&0.2509&0.7957&0.8284\\
\hline 
5&$I_{3322}$&0.0&0.25&0.2509&0.8177&0.8284\\
\hline 
6&$I_{3322}$&0.0&0.25&0.2509&0.7838&0.8284\\
\hline 
7&$I_{3322}$&0.0&0.25&0.2509&0.8147&0.8284\\
\hline 
8&$I_{3322}$&0.0&0.25&0.2509&0.7995&0.8284\\
\hline 
\end{longtable}

\end{center}
}

{

\setlength{\LTcapwidth}{\dimexpr\linewidth+4.5in\relax}

\begin{center}\setlength\tabcolsep{16pt} 

\begin{longtable}{ c c c c c c c  }

\caption[Properties of the Facets of the (3,3,4,2) scenario using ququarts ($d=4$).]{Properties of the Facets of the (3,3,4,2) scenario using ququarts ($d=4$).} \label{grid_A3342_4} \\
\hline \multicolumn{1}{c}{Nº} & \multicolumn{1}{c}{Name} & \multicolumn{1}{c}{$L$}& \multicolumn{1}{c}{$Q$}& \multicolumn{1}{c}{$Q_{NPA}$} & \multicolumn{1}{c}{$\lambda$} & \multicolumn{1}{c}{$\eta_{min}$}  \\ \hline 
\endfirsthead

\multicolumn{7}{c}%
{{\bfseries \tablename\ \thetable{} -- continued from previous page}} \\
\hline \multicolumn{1}{c}{Nº} & \multicolumn{1}{c}{Name} & \multicolumn{1}{c}{$L$}& \multicolumn{1}{c}{$Q$}& \multicolumn{1}{c}{$Q_{NPA}$} & \multicolumn{1}{c}{$\lambda$} & \multicolumn{1}{c}{$\eta_{min}$}   \\ \hline 
\endhead

\hline \multicolumn{7}{|r|}{{Continued on next page}} \\ \hline
\endfoot

\hline \hline
\endlastfoot
\hline
1&\text{CHSH}&0.0&0.2071&0.2071&0.7071&0.8284\\
\hline 
2&\text{CHSH}&0.0&0.2071&0.2071&0.7071&0.8284\\
\hline 
3&\text{CHSH}&0.0&0.2071&0.2071&0.7071&0.8284\\
\hline 
4&$I_{3322}$&0.0&0.25&0.2509&0.7758&0.8284\\
\hline 
5&$I_{3322}$&0.0&0.25&0.2509&0.8185&0.8284\\
\hline 
6&$I_{3322}$&0.0&0.25&0.2509&0.7755&0.8284\\
\hline 
7&$I_{3322}$&0.0&0.25&0.2509&0.819&0.8284\\
\hline 
8&$I_{3322}$&0.0&0.25&0.2509&0.799&0.8284\\
\hline 

\end{longtable}

\end{center}
}

\subsection{(2,2,4,4)}

{

\setlength{\LTcapwidth}{\dimexpr\linewidth+4.5in\relax}

\begin{center}\setlength\tabcolsep{16pt} 

\begin{longtable}{ c c c c c c c c }

\caption[Properties of the Facets of the (2,2,4,4) scenario using qubits ($d=2$).]{Properties of the Facets of the (2,2,4,4) scenario using qubits ($d=2$).} \label{grid_A2244_2} \\
\hline \multicolumn{1}{c}{Nº} & \multicolumn{1}{c}{Name} & \multicolumn{1}{c}{$L$}& \multicolumn{1}{c}{$Q$}& \multicolumn{1}{c}{$Q_{NPA}$} & \multicolumn{1}{c}{$\lambda$} & \multicolumn{1}{c}{$\eta_{min}$} & \multicolumn{1}{c}{C}  \\ \hline 
\endfirsthead

\multicolumn{8}{c}%
{{\bfseries \tablename\ \thetable{} -- continued from previous page}} \\
\hline \multicolumn{1}{c}{Nº} & \multicolumn{1}{c}{Name} & \multicolumn{1}{c}{$L$}& \multicolumn{1}{c}{$Q$}& \multicolumn{1}{c}{$Q_{NPA}$} & \multicolumn{1}{c}{$\lambda$} & \multicolumn{1}{c}{$\eta_{min}$} & \multicolumn{1}{c}{C}  \\ \hline 
\endhead

\hline \multicolumn{8}{|r|}{{Continued on next page}} \\ \hline
\endfoot

\hline \hline
\endlastfoot
\hline

1&$CHSH$&0.0&0.2071&0.2071&0.7071&0.8284&1.0\\
\hline 
2&$CHSH$&0.0&0.2071&0.2071&0.7071&0.8284&1.0\\
\hline 
3&$CHSH$&0.0&0.2071&0.2071&0.7071&0.8284&1.0\\
\hline 
4&$CHSH$&0.0&0.2071&0.2071&0.7071&0.8284&1.0\\
\hline 
5&$CHSH$&0.0&0.2071&0.2071&0.7071&0.8284&1.0\\
\hline 
6&$CHSH$&0.0&0.2071&0.2071&0.7071&0.8284&1.0\\
\hline 
7&$CHSH$&0.0&0.2071&0.2071&0.7071&0.8284&1.0\\
\hline 
8&$CHSH$&0.0&0.2071&0.2071&0.7071&0.8284&1.0\\
\hline 
\end{longtable}

\end{center}
}

{

\setlength{\LTcapwidth}{\dimexpr\linewidth+4.5in\relax}

\begin{center}\setlength\tabcolsep{16pt} 

\begin{longtable}{ c c c c c c c  }

\caption[Properties of the Facets of the (2,2,4,4) scenario using qutrits ($d=3$).]{Properties of the Facets of the (2,2,4,4) scenario using qutrits ($d=3$).} \label{grid_A2244_3} \\
\hline \multicolumn{1}{c}{Nº} & \multicolumn{1}{c}{Name} & \multicolumn{1}{c}{$L$}& \multicolumn{1}{c}{$Q$}& \multicolumn{1}{c}{$Q_{NPA}$} & \multicolumn{1}{c}{$\lambda$} & \multicolumn{1}{c}{$\eta_{min}$}   \\ \hline 
\endfirsthead

\multicolumn{7}{c}%
{{\bfseries \tablename\ \thetable{} -- continued from previous page}} \\
\hline \multicolumn{1}{c}{Nº} & \multicolumn{1}{c}{Name} & \multicolumn{1}{c}{$L$}& \multicolumn{1}{c}{$Q$}& \multicolumn{1}{c}{$Q_{NPA}$} & \multicolumn{1}{c}{$\lambda$} & \multicolumn{1}{c}{$\eta_{min}$}  \\ \hline 
\endhead

\hline \multicolumn{7}{|r|}{{Continued on next page}} \\ \hline
\endfoot

\hline \hline
\endlastfoot
\hline

1&$CHSH$&0.0&0.2071&0.2071&0.7071&0.8284\\
\hline 
2&$CHSH$&0.0&0.2071&0.2071&0.7137&0.8284\\
\hline 
3&$CHSH$&0.0&0.2071&0.2071&0.7002&0.8284\\
\hline 
4&$CHSH$&0.0&0.2071&0.2071&0.7137&0.8284\\
\hline 
5&$CHSH$&0.0&0.2071&0.2071&0.7002&0.8284\\
\hline 
6&$CHSH$&0.0&0.2071&0.2071&0.7071&0.8284\\
\hline 
7&$CHSH$&0.0&0.2071&0.2071&0.7105&0.8284\\
\hline 
8&$CHSH$&0.0&0.2071&0.2071&0.7037&0.8284\\
\hline 
\end{longtable}

\end{center}
}

{

\setlength{\LTcapwidth}{\dimexpr\linewidth+4.5in\relax}

\begin{center}\setlength\tabcolsep{16pt} 

\begin{longtable}{ c c c c c c c  }

\caption[Properties of the Facets of the (2,2,4,4) scenario using ququarts ($d=4$).]{Properties of the Facets of the (2,2,4,2) scenario using ququarts ($d=4$).} \label{grid_A2244_4} \\
\hline \multicolumn{1}{c}{Nº} & \multicolumn{1}{c}{Name} & \multicolumn{1}{c}{$L$}& \multicolumn{1}{c}{$Q$}& \multicolumn{1}{c}{$Q_{NPA}$} & \multicolumn{1}{c}{$\lambda$} & \multicolumn{1}{c}{$\eta_{min}$}  \\ \hline 
\endfirsthead

\multicolumn{7}{c}%
{{\bfseries \tablename\ \thetable{} -- continued from previous page}} \\
\hline \multicolumn{1}{c}{Nº} & \multicolumn{1}{c}{Name} & \multicolumn{1}{c}{$L$}& \multicolumn{1}{c}{$Q$}& \multicolumn{1}{c}{$Q_{NPA}$} & \multicolumn{1}{c}{$\lambda$} & \multicolumn{1}{c}{$\eta_{min}$}   \\ \hline 
\endhead

\hline \multicolumn{7}{|r|}{{Continued on next page}} \\ \hline
\endfoot

\hline \hline
\endlastfoot
\hline

1&$CHSH$&0.0&0.2071&0.2071&0.7071&0.8284\\
\hline 
2&$CHSH$&0.0&0.2071&0.2071&0.7072&0.8284\\
\hline 
3&$CHSH$&0.0&0.2071&0.2071&0.6993&0.8284\\
\hline 
4&$CHSH$&0.0&0.2071&0.2071&0.7208&0.8284\\
\hline 
5&$CHSH$&0.0&0.2071&0.2071&0.6978&0.8284\\
\hline 
6&$CHSH$&0.0&0.2071&0.2071&0.7071&0.8284\\
\hline 
7&$CHSH$&0.0&0.2071&0.2071&0.7141&0.8284\\
\hline 
8&$CHSH$&0.0&0.2071&0.2071&0.7106&0.8284\\
\hline 

\end{longtable}

\end{center}
}

\subsection{(4,3,3,2)}

{

\setlength{\LTcapwidth}{\dimexpr\linewidth+4.5in\relax}

\begin{center}\setlength\tabcolsep{16pt} 

\begin{longtable}{ c c c c c c c c }

\caption[Properties of the Facets of the (4,3,3,2) scenario using qubits ($d=2$).]{Properties of the Facets of the (4,3,3,2) scenario using qubits ($d=2$).} \label{grid_A4332_2} \\
\hline \multicolumn{1}{c}{Nº} & \multicolumn{1}{c}{Name} & \multicolumn{1}{c}{$L$}& \multicolumn{1}{c}{$Q$}& \multicolumn{1}{c}{$Q_{NPA}$} & \multicolumn{1}{c}{$\lambda$} & \multicolumn{1}{c}{$\eta_{min}$} & \multicolumn{1}{c}{C}  \\ \hline 
\endfirsthead

\multicolumn{8}{c}%
{{\bfseries \tablename\ \thetable{} -- continued from previous page}} \\
\hline \multicolumn{1}{c}{Nº} & \multicolumn{1}{c}{Name} & \multicolumn{1}{c}{$L$}& \multicolumn{1}{c}{$Q$}& \multicolumn{1}{c}{$Q_{NPA}$} & \multicolumn{1}{c}{$\lambda$} & \multicolumn{1}{c}{$\eta_{min}$} & \multicolumn{1}{c}{C}  \\ \hline 
\endhead

\hline \multicolumn{8}{|r|}{{Continued on next page}} \\ \hline
\endfoot

\hline \hline
\endlastfoot
\hline
1&\text{CHSH}&0.0&0.2071&0.2071&0.7071&0.8284&1.0\\
\hline 
2&$I_{3322}$&0.0&0.25&0.2509&0.8&0.8284&1.0\\
\hline 
3&$I_{3322}$&0.0&0.25&0.2509&0.8&0.8284&1.0\\
\hline 
4&$I_{3322}$&0.0&0.25&0.2509&0.8&0.8284&1.0\\
\hline 
5&$I_{3322}$&1.0&1.4142&1.4142&0.7071&0.8284&1.0\\
\hline 
6&$I_{4322}^1$&0.0&0.4142&0.4142&0.7071&0.8284&1.0\\
\hline 
7&$I_{4322}^1$&1.0&1.4142&1.4142&0.7071&0.8284&1.0\\
\hline 
8&$I_{4322}^2$&0.0&0.299&0.299&0.7698&0.8699&1.0\\
\hline 
\end{longtable}

\end{center}
}

{

\setlength{\LTcapwidth}{\dimexpr\linewidth+4.5in\relax}

\begin{center}\setlength\tabcolsep{16pt} 

\begin{longtable}{ c c c c c c c  }

\caption[Properties of the Facets of the (4,3,3,2) scenario using qutrits ($d=3$).]{Properties of the Facets of the (4,3,3,2) scenario using qutrits ($d=3$).} \label{grid_A4332_3} \\
\hline \multicolumn{1}{c}{Nº} & \multicolumn{1}{c}{Name} & \multicolumn{1}{c}{$L$}& \multicolumn{1}{c}{$Q$}& \multicolumn{1}{c}{$Q_{NPA}$} & \multicolumn{1}{c}{$\lambda$} & \multicolumn{1}{c}{$\eta_{min}$}   \\ \hline 
\endfirsthead

\multicolumn{7}{c}%
{{\bfseries \tablename\ \thetable{} -- continued from previous page}} \\
\hline \multicolumn{1}{c}{Nº} & \multicolumn{1}{c}{Name} & \multicolumn{1}{c}{$L$}& \multicolumn{1}{c}{$Q$}& \multicolumn{1}{c}{$Q_{NPA}$} & \multicolumn{1}{c}{$\lambda$} & \multicolumn{1}{c}{$\eta_{min}$}  \\ \hline 
\endhead

\hline \multicolumn{7}{|r|}{{Continued on next page}} \\ \hline
\endfoot

\hline \hline
\endlastfoot
\hline
1&\text{CHSH}&0.0&0.2071&0.2071&0.7071&0.8284\\
\hline 
2&$I_{3322}$&0.0&0.25&0.2509&0.8067&0.8284\\
\hline 
3&$I_{3322}$&0.0&0.25&0.2509&0.8095&0.8284\\
\hline 
4&$I_{3322}$&0.0&0.25&0.2509&0.7991&0.8284\\
\hline 
5&$I_{3322}$&1.0&1.4142&1.4142&0.7365&0.8284\\
\hline 
6&$I_{4322}^1$&0.0&0.4142&0.4142&0.7529&0.8284\\
\hline 
7&$I_{4322}^1$&1.0&1.4142&1.4142&0.7449&0.8284\\
\hline 
8&$I_{4322}^2$&0.0&0.299&0.299&0.7825&0.8699\\
\hline

\end{longtable}

\end{center}
}

{

\setlength{\LTcapwidth}{\dimexpr\linewidth+4.5in\relax}

\begin{center}\setlength\tabcolsep{16pt} 

\begin{longtable}{ c c c c c c c  }

\caption[Properties of the Facets of the (4,3,3,2) scenario using ququarts ($d=4$).]{Properties of the Facets of the (4,3,3,2) scenario using ququarts ($d=4$).} \label{grid_A4332_4} \\
\hline \multicolumn{1}{c}{Nº} & \multicolumn{1}{c}{Name} & \multicolumn{1}{c}{$L$}& \multicolumn{1}{c}{$Q$}& \multicolumn{1}{c}{$Q_{NPA}$} & \multicolumn{1}{c}{$\lambda$} & \multicolumn{1}{c}{$\eta_{min}$}  \\ \hline 
\endfirsthead

\multicolumn{7}{c}%
{{\bfseries \tablename\ \thetable{} -- continued from previous page}} \\
\hline \multicolumn{1}{c}{Nº} & \multicolumn{1}{c}{Name} & \multicolumn{1}{c}{$L$}& \multicolumn{1}{c}{$Q$}& \multicolumn{1}{c}{$Q_{NPA}$} & \multicolumn{1}{c}{$\lambda$} & \multicolumn{1}{c}{$\eta_{min}$}   \\ \hline 
\endhead

\hline \multicolumn{7}{|r|}{{Continued on next page}} \\ \hline
\endfoot

\hline \hline
\endlastfoot
\hline
1&\text{CHSH}&0.0&0.2071&0.2071&0.7071&0.8284\\
\hline 
2&$I_{3322}$&0.0&0.25&0.2509&0.7829&0.8284\\
\hline 
3&$I_{3322}$&0.0&0.25&0.2509&0.8128&0.8284\\
\hline 
4&$I_{3322}$&0.0&0.25&0.2509&0.799&0.8284\\
\hline 
5&$I_{3322}$&1.0&1.4142&1.4142&0.7392&0.8284\\
\hline 
6&$I_{4322}^1$&0.0&0.4142&0.4142&0.7695&0.8284\\
\hline 
7&$I_{4322}^1$&1.0&1.4142&1.4142&0.7553&0.8284\\
\hline 
8&$I_{4322}^2$&0.0&0.299&0.299&0.7861&0.8699\\
\hline 

\end{longtable}

\end{center}
}

\end{document}